\documentclass[journal, magazine]{IEEEtran}
\usepackage{amsmath,amsfonts}
\usepackage{algorithmic}
\usepackage{algorithm}
\usepackage{array}
\usepackage[caption=false,font=normalsize,labelfont=sf,textfont=sf]{subfig}
\usepackage{textcomp}
\usepackage{stfloats}
\usepackage{url}
\usepackage{verbatim}
\usepackage{graphicx}
\usepackage{cite}
\usepackage{epstopdf}
\usepackage{epsfig}
\usepackage{color}
\usepackage{makecell}

\begin{document}
	
	\title{Semantic Communication for Edge Intelligence\\ Enabled Autonomous Driving System}
	
	\author{
		Yunqi~Feng,
		Hesheng~Shen,
		Zhendong~Shan,
		Qianqian~Yang,~\IEEEmembership{Member,~IEEE},
		and Xiufang~Shi,~\IEEEmembership{Member,~IEEE}
		
		\thanks{This work was supported in part by the National Natural Science Foundation of China under Grant 62301491, and Grant U23A20326, in part by the Zhejiang Provincial Natural Science Foundation of China under Grant LQ23F010023, \emph{(Corresponding author: Xiufang~Shi).}}
		
		\thanks{Yunqi Feng, Hesheng Shen, Zhendong Shan and Xiufang Shi are with the College of Information Engineering, Zhejiang University of Technology, Hangzhou, China. (e-mail:yqfeng@zjut.edu.cn, 221122030366@zjut.edu.cn, zhendongshan@zjut.edu.cn, xiufangshi@zjut.edu.cn)}
		\thanks{Qianqian Yang is with the College of Information Science and Electronic Engineering, Zhejiang University, Hangzhou, China. (e-mail: qianqianyang20@zju.edu.cn)}
	}

	\markboth{}%
	{}

	\maketitle

	\begin{abstract}
		Expected to provide higher transportation efficiency and security, autonomous driving has attracted substantial attentions from both industry and academia. Meanwhile, the emergence of edge intelligence has further introduced significant advancements to this field.
		However, the crucial demands of ultra-reliable and low-latency communications (URLLC) among the vehicles and edge servers have hindered the development of autonomous driving.
		In this article, we provide a brief overview of edge intelligence enabled autonomous driving system and current vehicle-to-everything (V2X) technologies. Moreover, challenges associated with massive data transmission in autonomous driving are highlighted from three perspectives: multi-modal data transmission and fusion, multi-user collaboration and connection, and multi-task training and execution.
		To cope with these challenges, we propose to incorporate semantic communication into autonomous driving to achieve highly efficient and task-oriented data transmission. Unlike traditional communications, semantic communication extracts task-relevant semantic feature from multi-sensory data. Specifically, a unified multi-user semantic communication system for transmitting multi-modal data and performing multi-task execution is designed for collaborative data transmission and decision making in autonomous driving.
		Simulation results demonstrate that the proposed system can significantly reduce data transmission volume without compromising task performance, as evidenced by the realization of a cooperative multi-vehicle target classification and detection task.
	\end{abstract}
	
	
	\section{Introduction}
	Autonomous driving has gained increasing attentions due to its significant advantages in mitigating traffic accidents and alleviating congestion. The internet of vehicles (IoV), relying on vehicle-to-everything (V2X) technologies, is the crucial foundation of autonomous driving\cite{ref1}. With the IoV, autonomous vehicles can access sufficient information to make accurate decisions, while passengers can enjoy various emerging on-vehicle services. However, single-vehicle approaches are limited by constrained computing power and storage capacity. Meanwhile, transmitting massive data to a centralized cloud computing center will lead to high transmission latency and significant pressure on the backbone network. Consequently, edge intelligence-enabled autonomous driving has been widely studied because of its capability in facilitating lower latency and reduced backbone network pressure\cite{ref2}.
	
	To meet the requirement of ultra-reliable and low-latency communications (URLLC) in autonomous driving, massive multi-modal data need be transmitted among various devices within limited delay and integrity restrictions, leading to a huge demand for spectrum resources\cite{ref1}. In traditional communication systems, such as 802.11p-based V2X and cellular V2X (C-V2X), all source data collected by the vehicles are transmitted. Although the latest advancements of traditional communication have been integrated into the V2X to improve spectrum efficiency, the available spectrum resources become constrained as the number of connected vehicles and the complexity of intelligent tasks continue to increase\cite{ref3}. Moreover, the high mobility of vehicles, as well as the heterogeneity and modality-specificity of communication technologies, deteriorate the spectrum efficiency and exacerbate the complexity of data fusion\cite{ref4}.
	
	With a high degree of flexibility and adaptivity, semantic communication has been proposed as one of the potential technologies for the future 6G era\cite{ref5}. Unlike traditional communication, semantic communication extracts semantic features and minimizes the transmission of redundant data while maintaining communication effectiveness\cite{ref6}. Since the perception tasks serve as the foundation of autonomous driving~\cite{ref10}, the reliance on deep learning (DL) for high-dimensional feature extraction promotes the combination of semantic communication and edge intelligence-enabled autonomous driving.
	Moreover, by constructing a semantic knowledge base, semantic communication is expected to address the incompatibilities caused by inconsistent information modalities\cite{ref7}. It will facilitate alignment, fusion, and decision-making of multi-modal information.
	
	Although there have been extensive researches regarding semantic communication\cite{ref5, ref6, ref7}, few semantic communication systems have been designed specifically for autonomous driving. The authors in \cite{ref8} developed a semantic communication system for the reconstruction of image segmentation in the IoV. However, only the single-link vehicle-to-vehicle image communication scenario was considered, and collaborative sensing and decision-making for multiple vehicles were not discussed. In\cite{ref9}, the authors focused solely on the multi-user approach without considering different data modalities and correlations between various tasks. Considering the relevance and redundancy of multi-modal data from diverse sources, effective collaboration mechanisms should be considered. Moreover, coordinating the execution process of multiple tasks based on their interdependence can further enhance efficiency.

	In this work, we propose a unified multi-user semantic communication system for transmitting multi-modal data and executing multiple tasks in autonomous driving.
	First, we jointly design semantic and channel coding and fusion modules for different users and data modalities, which not only significantly reduces the transmission redundancy of multi-modal data, but also improves the integration efficiency of the edge server. Additionally, the design of the task performer takes into account multiple tasks in different levels, enabling adaptive task execution at either the feature level or the data level based on the specific requirements of each task. Moreover, the task performer can also coordinate the execution order and process, which maximizes the utilization of continuity and correlation between tasks. Finally, the advantages of the proposed system are highlighted by implementing a case study. In summary, the proposed system has the capability to efficiently transmit multi-modal data related to multi-tasks with limited communication resources. Additionally, the server is capable of aggregating data from multiple users and coordinating multi-task decisions, thus facilitating secure and reliable multi-vehicle collaboration.

	\begin{figure*}[!t]
		\centering
		\includegraphics[width=6in]{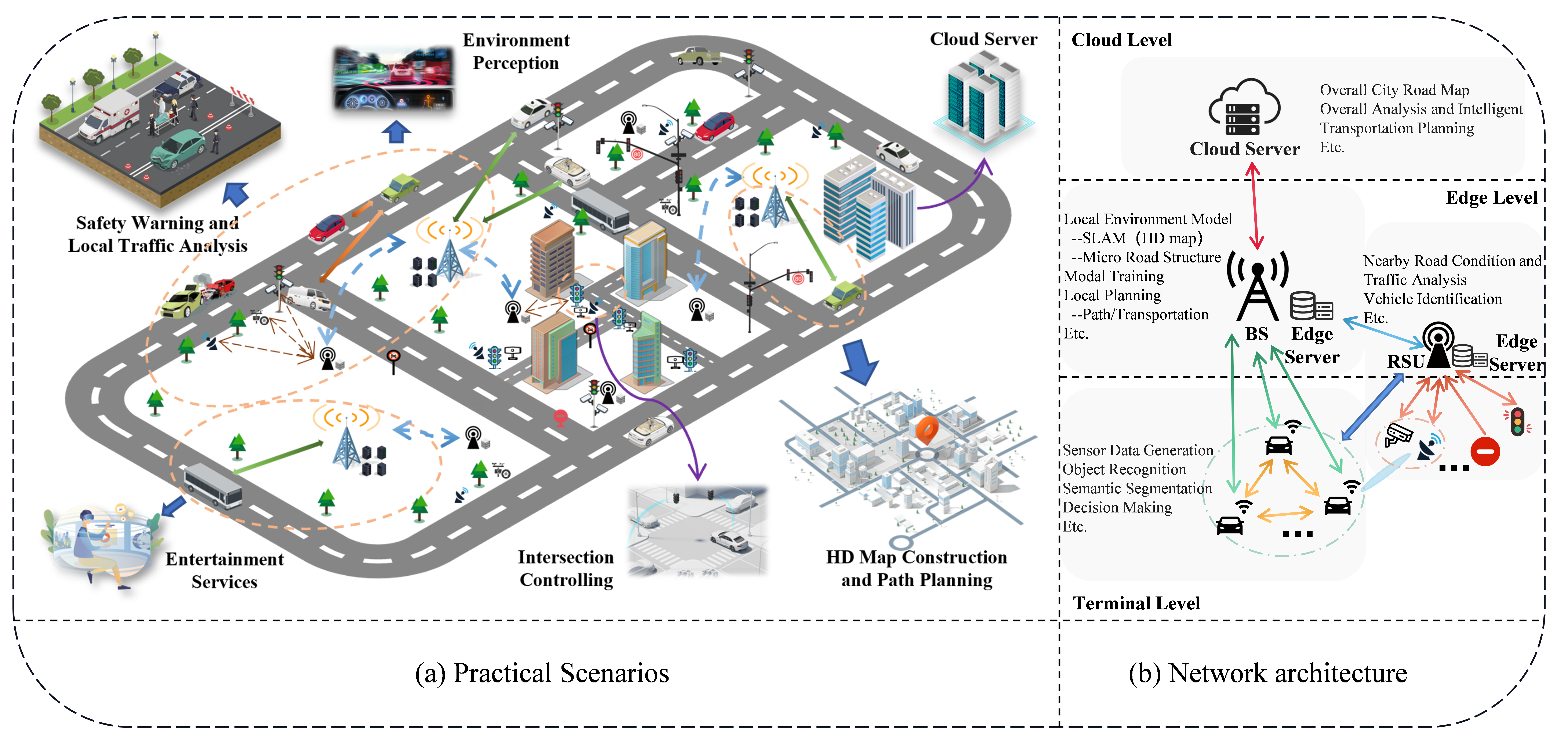}
		\caption{Practical scenarios and network architecture of edge intelligence enabled autonomous driving system.}
		\label{fig_1}
	\end{figure*}
	
	\section{Use Cases, Current Advances and Communication Challenges}
	
	In this section, we provide a brief overview of the use cases and current advances in the V2X. Subsequently, three main challenges in the edge intelligence enabled autonomous driving system are concluded.
	
	\subsection{Use Cases and Applications}
	
	As shown in Fig. \ref {fig_1} (a), this article summaries the practical scenarios of edge intelligence enabled autonomous driving system, encompassing a range of real-time applications \cite{ref1, ref3, ref4}, which can be classified into three groups:
	\begin{itemize}
		\item Safety: This group covers basic safety use cases such as collision warning, as well as more advanced ones like intersection controlling.
	\end{itemize}
	\begin{itemize}
		\item Extended Sensors: This group enables the exchange of sensory data among vehicles, road side units (RSU), edge servers, and cloud servers, including environment perception and high-definition (HD) map construction.
	\end{itemize}
	\begin{itemize}
		\item Convenience: This group encompasses use cases that deliver value and convenience to either the driver or the management company, such as entertainment services.
	\end{itemize}
	
	Fig. \ref {fig_1} (b) presents the network architecture of edge intelligence enabled autonomous driving system\cite{ref2}, which consists of the terminal level, the edge level, and the cloud level. The terminal level comprises intelligent devices and autonomous vehicles, which are responsible for collecting road and environmental information. Afterwards, the devices and vehicles upload the collected data to the edge level for further integration and utilization. Moreover, intelligent devices can assist in traffic management, while autonomous vehicles can exchange information with nearby vehicles and devices. The cloud level consisting of clustered servers, serves as a central hub for aggregating data from diverse regions.
	
	\subsection{Current Advances in the V2X}
	
	In the 802.11p-based V2X,  dedicated short-range communication (DSRC) is utilized to support low-latency communication, enabling vehicles to promptly share critical information such as road conditions and vehicle status. However, DSRC has some limitations, such as line-of-sight communication capability, which renders vehicles susceptible to intermittent connectivity when moving at high speeds\cite{ref4}. Additionally, DSCR primarily transmits small awareness messages, which is not suitable for multi-modal visual data\cite{ref1}. By contrast, C-V2X brings higher data transmission rate, lower latency and better connection density. However, the growing number of vehicles in autonomous driving lead to spectrum congestion in cellular networks\cite{ref4}. With regards to multi-user collaboration, it is challenging to establish a reliable resource coordination mechanism for collaborative data exchange and connectivity among multiple users. Additionally, the current task execution and data transmission in the V2X is often designed separately, which results in the transmission of redundant data from various sources with multi-modal and may lead to repetitive requests for the same data from similar tasks.
	
	\subsection{Communication Challenges}
	
	Based on previous research \cite{ref1, ref2, ref3, ref4, ref10}, three main communication challenges in edge intelligence enabled autonomous driving can be concluded:
	
	\subsubsection{Multi-modal Data Transmission and Fusion}
	
	In autonomous driving, the expansion of data modalities has encompassed a wide range of types, including images, point clouds and radar signals. The utilization of multi-modal data can provide valuable information for autonomous driving and improve decision-making accuracy. However, the transmitted data often contain a significant amount of redundant information, which creates substantial communication pressure on autonomous driving. Moreover, the lack of unified representation for multi-modal data further induces extra processing time for data-level fusion before specific tasks can be performed. Therefore, it is essential to explore techniques that can extract useful multi-modal information collectively with minimum redundancy, and simplify the data fusion process at the receiving end.

	\subsubsection{Multi-user Collaboration and Connection}
	
	The growing number of connected vehicles in autonomous driving necessitates efficient multi-user collaboration. For example, the construction of the environment model relies on multiple devices providing perception data simultaneously. Consequently, effective collaboration mechanisms are required to make full use of the distributed data and optimize resource allocation. Moreover, the abundance of data sources emphasizes the necessity of establishing effective multi-user connections among these sources. Despite the existence of numerous protocols, interfaces, and emerging cross-protocol communication technologies, transmitting data across different networks, protocols, and transmission technologies remains a challenge. Obviously, a straightforward and universally applicable technology is required for achieving multi-user connections across networks and protocols for the cross-network and cross-protocol wireless communication.
	
	\subsubsection{Multi-task Joint Training and Execution}
	
	A significant characteristic of autonomous driving is the interdependence among various tasks. Multiple tasks may depend on the same data, and the result of one task may influence other tasks. For example, the data used in object detection can also be applied to area segmentation and lane detection, and the object detection is involved in assisting vehicles with obstacle avoidance and adherence to traffic regulations. Although numerous methods have been developed in performing various tasks separately in autonomous driving, it often results in excessive time and computational resource consumption. Furthermore, providing the same data to multiple tasks significantly increases communication overhead. Therefore, exploring multi-task joint training and execution can reduce the computation and communication overhead.
	
	\section{Unified Multi-user Semantic Communication System for Autonomous Driving}
	
\begin{figure*}[!t]
		\centering
		\includegraphics[width=6in]{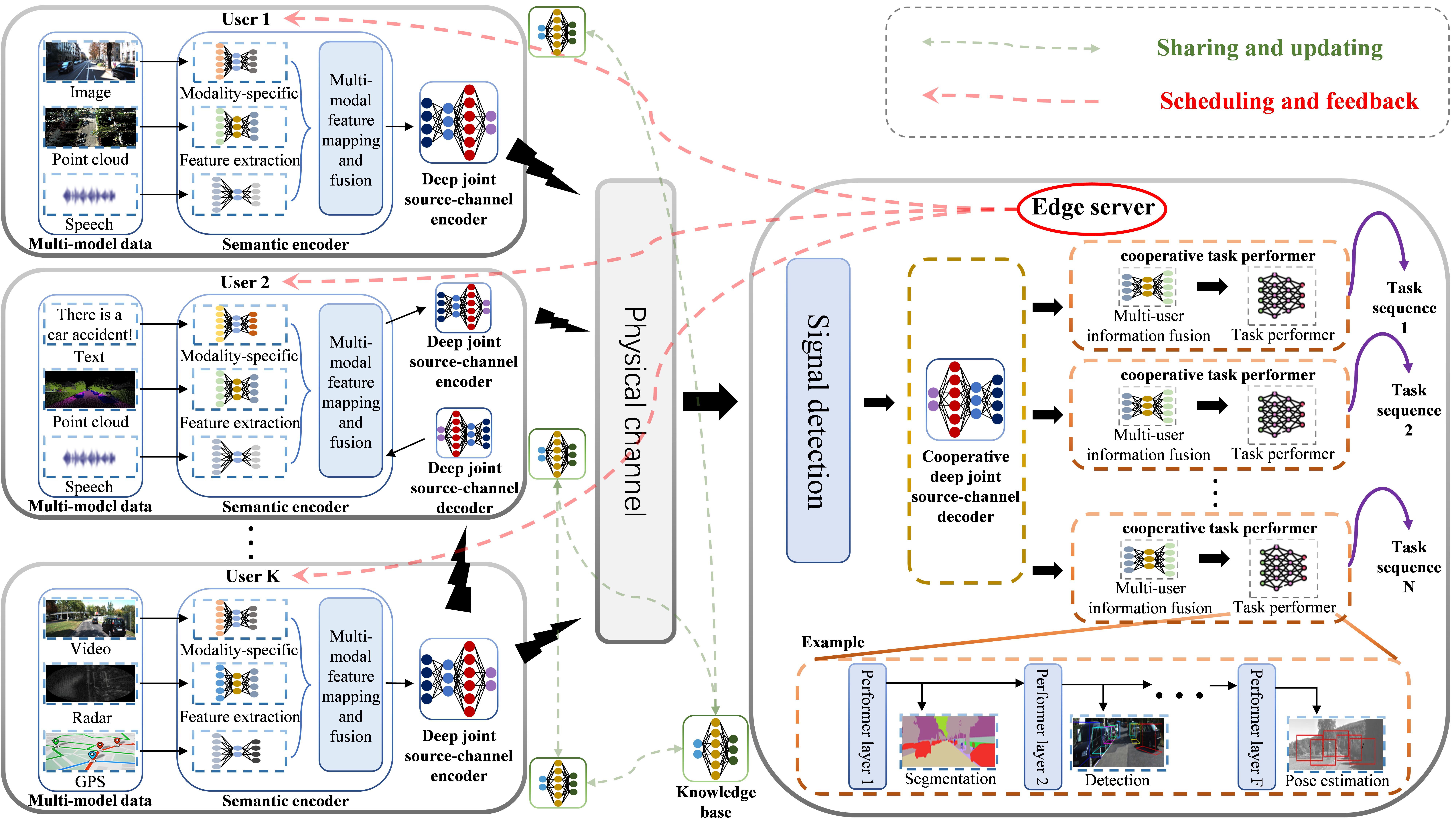}
		\caption{Unified multi-user semantic communication for autonomous driving.}
		\label{fig_2}
	\end{figure*}

	In this section, we proposed a unified multi-user semantic communication system for edge intelligence enabled autonomous driving. As illustrated in Fig. \ref {fig_2}, the system consists of the semantic encoder, deep joint source-channel encoder/decoder (JSC-Codec), signal detection module and cooperative task performers. The semantic features extracted in the semantic encoder are further compressed and recovered by the JSC-Codec with the same knowledge base. To reduce redundant data transmissions, the semantic encoder and the cooperative task performers are jointly optimized, where the semantic-level correlation among different users can be learned. Then, the semantic-driven task performers can successfully execute a series of tasks. Furthermore, the construction of knowledge bases ensures consistent understanding of the semantic information across all participating devices, thereby enhancing the reliability of transmission.
	
	\subsection{Semantic Encoder}
	
	The semantic encoder is designed to extract semantic feature from multi-modal data and fuse them into a unified high-dimensional representation, encompassing task-relevant information of the source data. The architecture of the semantic encoder depends on the modalities of source data and the tasks being executed\cite{ref5}. For instance, convolutional neural networks (CNNs) or vision transformers are commonly employed for image processing tasks, while graph neural networks (GNNs) are preferred for three-dimensional object/scene models. Apart from GNN, PointNet is also used to handle point cloud data, and long short-term memory (LSTM) networks are applied to effectively process time series data.
	
	The semantic features extraction relies on existing knowledge base, such as HD maps, environment models and the type of specific task. For example, when updating HD maps, only the latest updates that differ from the original map, such as preceding traffic and pedestrians on the roadside need to be extracted. Moreover, since the multi-modal data of the same user often possess redundancy, it is crucial to map the semantic features to a unified semantic space\cite{ref7}, where semantic features that belong to the similar information are closer together. Specifically, the process of feature alignment and fusion occurs in the multi-modal feature mapping and fusion module, incorporating contrast learning techniques. Additionally, inter-modal complementarity learning can be conducted to effectively enhance the quality and task-specificity of transmitted data, facilitating subsequent feature fusion processes.

	\subsection{JSC-Codec}
	
	The deep joint source and channel coding (JSCC) is widely adopted in semantic communication to combat channel distortions and ensure the efficient delivery of semantic information~\cite{ref6}.
	The JSC encoder utilizes the deep neural networks to encode the semantic information as the transmitted symbols. Since transmitted symbols are obtained at the semantic level, the importance level of the semantic information can be learned. Hence, the JSC encoder utilizes different numbers of transmitted symbols to encode semantic information with different importance level and demonstrates robustness to the channel distortions, which can avoid the ``cliff effect" performs admirably at low signal-to-noise ratio (SNR)~\cite{ref5}.
	
	At the receiver, the semantic information from multiple users can be recovered by the JSC decoder. Moreover, the semantic-level correlation between multiple users can be exploited to improve the accuracy of the cooperative JSC decoder based on the similarity of the received symbols of different users. In addition, to support various tasks with multi-modal data in multi-user semantic communication, the collaborative JSC decoder further merges the semantic information from different users. Specifically, unlike conventional semantic communications, the proposed system employs multiples user to serve multiple tasks. For example, the single-user data are often limited to a specific perspective of the current driving scene, hindering the acquisition of comprehensive scene information. Collecting multi-user data from various sensors allows the edge server to obtain the semantic information in a complementary manner.

	\subsection{Cooperative Task Performers}
	
	The cooperative task performers play a crucial role in accomplishing specific sequential tasks with the recovered semantic information from multiple users\cite{ref8}. In the proposed system, since some intelligent tasks involve machine-to-machine applications, semantic information can be directly used by the cooperative task performers. Specifically, cooperative task performers can adaptively execute at the semantic level or the data level based on the type of task. The direct use of semantic features avoids the process of source data reconstruction, significantly reducing the computational complexity of tasks such as target classification. Moreover, the source data can restore the original scene, which is suitable for target detection, semantic segmentation and other complex tasks. In addition, in order to improve the performance of the cooperative task performers, the correlation and complementary of the semantic information among multiple users can be fully utilized. The multi-user information fusion module can facilitate better task performance by combining the semantic information transmitted from multiple users with task-specific combination scheme, which is adaptively designed based on task types and data modalities.
	
	Meanwhile, inspired by the inherent relevance of intelligent tasks in autonomous driving, the proposed system incorporates multi-task joint training and execution to leverage multi-user data more effectively and improve the task execution performance. The cooperative task performers learn the correlations between different intelligent tasks, enabling each task performer to execute a series of tasks based on the same semantic information and mutual guidance from other task performers. For example, in the tasks like segmentation, detection, and pose estimation, the task execution process follows a step-by-step approach. At each step, the task performer combines the correlated semantic information with the results of previous tasks, creating an enhanced semantic feature that results in improved task performance.

	\subsection{Knowledge Base}

    The premise of semantic communication is the shared knowledge base\cite{ref5,ref6}, which contains background knowledge and enables the vehicles and edge servers have the same understanding of semantic information. With the knowledge base, the system can extract and fuse multi-user semantic features across various model architectures, ensuring the safe and effective operations of autonomous vehicles. Additionally, in the proposed system, the limitations of computing and storage resources, as well as the need for privacy protection in the vehicles are further considered. As a result, the construction and updates of the shared knowledge base are performed by edge servers, while vehicles exclusively update their own private knowledge base.

	\subsubsection{Shared Knowledge Base and Private Knowledge Base}

    The vehicles serve as exclusive terminals for executing autonomous driving tasks with fundamental similarity in task content. Consequently, a significant portion of the knowledge, including city maps and feature extraction techniques, can be shared among these vehicles. The shared knowledge base encompasses essential background information for autonomous driving tasks and is stored in the edge server to ensure accessibility. However, since each vehicle also possesses distinct vehicle-specific information, constructing a private knowledge base for each vehicle is necessary. Moreover, to facilitate continuous communication between the vehicles and edge servers, the private knowledge base of each vehicle can be rapidly propagated to subsequent edge servers based on the recent itinerary plan. Given that the shared knowledge base encompasses most of the background knowledge, the private knowledge base tends to be relatively small, resulting in minimal resource occupation during the transmission process.

	\subsubsection{Knowledge Bases Updating}
	
	Given the dynamic nature of traffic environments in autonomous driving, regular knowledge base updates are imperative. By continuously refining and updating the knowledge base, we can significantly enhance the semantic understanding of the autonomous driving system. This, in turn, reduces the likelihood of ambiguities and improves the overall reliability and safety of autonomous driving. The updates of private knowledge base are exclusively handled by the vehicle itself, whereas the shared knowledge base allows distributed updates from all RSUs and online connected vehicles. Specifically, to maintain an up-to-date shared knowledge base, RSUs and vehicles promptly upload knowledge updates, which are then aggregated by edge servers with the original knowledge base to obtain the latest version. The updated knowledge is subsequently disseminated to other edge servers and surrounding vehicles, thus completing the devices-servers-devices update process.

	\subsection{Model Training and Channel Conditions}

	In the proposed system, various models including autonomous driving models and semantic feature extraction models need to be trained or updated. The initialization and subsequent updates of the models can be performed by edge servers with the shared knowledge base, which includes vehicle-collected datasets consisting of traffic and communication statuses. Once the model parameters are acquired, each vehicle assesses the model with its own data, labels incorrect data pairs, and provides feedback on the evaluation results and new annotation data to the edge servers. The edge servers further optimize the model by aggregating evaluation results and annotation data from multiple vehicles, subsequently broadcasting the refined model. This iterative process continues until the models achieve high accuracy and generalization.
	In addition to the cross-level model training, edge servers can also update models through devices at the same level. Since edge servers are typically deployed in specific locations and training data often come from vehicles and RSUs in the nearby region, resulting in models with strong regional characteristics. To improve model generalization ability, federated learning can be used to exchange model parameters between different edge servers, facilitating the training and aggregation of models from different regions\cite{ref4}. Moreover, to enhance the transmission performance of the proposed system, which primarily relies on physical layer communication technologies, it is essential to model the physical channel as a series of non-trainable layers during training. Furthermore, advanced DL-based channel estimation and equalization technologies can be leveraged to mitigate the impact of physical channels\cite{ref6}.
	
	\begin{figure*}[!t]
		\centering
		\includegraphics[width=6in]{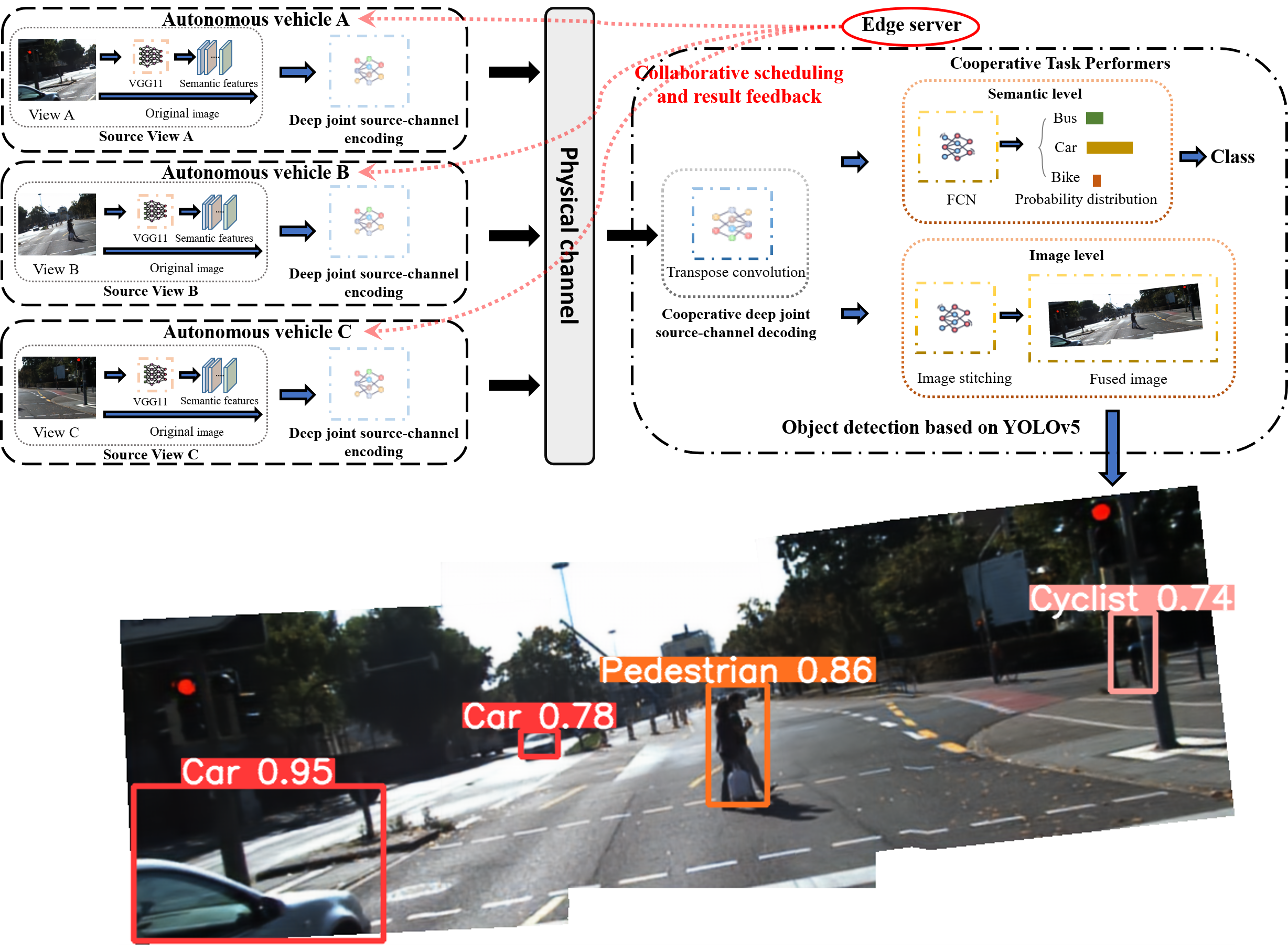}
		\caption{The framework for cooperative multi-vehicle semantic communication (COMV-SC).}
		\label{fig_3}
	\end{figure*}
	
	\section{Case Study: Cooperative Multi-Vehicle Semantic Communication}

    In this section, we present a case study to demonstrate the effectiveness of the proposed system. Specifically, a multi-modal multi-task framework for cooperative multi-vehicle semantic communication (COMV-SC) is proposed to improve the perceptual performance of autonomous vehicles. Since the decision making and instruction execution of vehicles heavily rely on comprehensive and accurate target classification and detection results, each vehicle needs to receive and aggregate target images captured from surrounding vehicles with multiple views to achieve a wider field of perspective. However, this approach may lead to redundant transmissions and necessitate additional computational power and communication bandwidth in vehicles. Hence, in the COMV-SC, only task-relevant semantic features are transmitted from each vehicle to the edge server, where target classification and detection can be performed at multiple levels by fusing information from multiple users, as illustrated in Fig. \ref {fig_3}. Meanwhile, a random user is designated, and the thermal modality is incorporated to achieve robust and accurate semantic segmentation, especially under unsatisfied lighting conditions.
	
	\subsection{Methodology}

    At the transmitter, the semantic features of target images are first extracted and encoded with the JSCC network. The pretrained JSCC network enables consistent image transmission quality across various open datasets with varying data volumes and sizes, facilitates the proposed system with effectiveness and generality. For tasks that can be executed at the semantic level such as target classification, the CNN network of VGG-11\cite{ref11} is used to extract the per-user view semantic information before going through the JSCC. Moreover, as a key component of the proposed system, the multi-modal feature mapping and fusion module is composed of two modified ResNet50 networks linked by several short cuts. At the receiver, after signal detection, the cooperative JSC decoder jointly decodes the detected symbols to restore the semantic information or target images of multiple users. According to the type of the task, the cooperative task performers adaptively execute the task at the semantic level or the data level, respectively. For tasks such as target classification, the cooperative task performers fuse the multi-user semantic features at the semantic level and then feed them into the fully connected network (FCN) to get the classification results. For tasks such as target detection, where source images need to be recovered from the high-dimensional semantic information, the cooperative task performers can stitch multiple target images to create a comprehensive wide-perspective image with the parallax-tolerant image stitching method in\cite{ref12}. Finally, the You Only Look Once v5s (YOLOv5s) algorithm is adopted to identify and label the target location.
	
	\subsection{Experiment Setups}
	
	\subsubsection{Datasets}

    We train the JSCC network, the classification network, and the YOLOv5s network (fine-tuned) with the mini-Imagenet dataset\cite{ref15}, the multi-view 3D dataset\cite{ref11} and the KITTI dataset\cite{ref14}, respectively. Moreover, due to the shortage of datasets relevant to image stitching, we randomly cropped each original image in the testing set of the KITTI dataset into three 400 × 300 images. During image cropping, each image is subjected to a random rotation in the range of -10 to 10 degrees, in addition to a 20\% to 50\% overlap of two adjacent images. To simplify the experiment, the ``DontCare" and ``Misc" categories are excluded. As for multi-modal semantic segmentation, we have employed the RGB-Thermal dataset in~\cite{ref13}.
	
	\subsubsection{Evaluation Metrics}
	
    For the classification task, precision and recall are two important metrics that can assess the impact of false negatives and false positives, respectively. When both the metrics are considered, the F1 score is employed since it represents the harmonic mean between precision and recall and offers a comprehensive evaluation. Since the F1 score is class-specific, we calculate the weighted mean across all classes to obtain a generalized score. And for the detection task, the mAP50 (mean Average Precision at an IoU threshold of 0.50) metric is adopted to represent the accuracy of target detection. Similarly, the average class accuracy (class avg.) and the mean intersection over union (mean IoU) are employed to evaluate the semantic segmentation performance.

	\subsubsection{Training Strategy}
	The proposed system employs a two-stage training method. In the first stage, the JSCC network, the classification network, and YOLOv5s network are trained separately. In the second stage, different networks are combined together to form the overall COMV-SC network, which is further fine-tuned end-to-end to improve the overall performance. Moreover, the JSCC network is trained under an additive white Gaussian noise (AWGN) channel with ${\rm SNR} = 0$.
	
	\subsubsection{Compared Methods}
	
    The proposed COMV-SC is compared with traditional method, employing the JPEG format with 30\% compression ratio for source encoding and LDPC with a 1/2 rate for channel encoding, and QPSK for modulation, which is referred to as `JPEG+LDPC+QPSK’. After recovering the image at the receiver, target classification is performed using VGG-11 and target detection is performed using YOLOv5s, respectively. For the classification task, we also include the voting scheme, where each vehicle classifies the target based on the captured single-view image and then transmits the result to the edge server for acquiring the final classification result. For the COMV-SC, we use the JSCC network with a compression rate of 8.4\% in the classification task and 16.7\% in the target detection task. Additionally, the MFNet and RTFNet without transmission~\cite{ref13} have been chosen for  comparison in multi-modal semantic segmentation.

	\begin{figure}[!t]
		\centering
		\includegraphics[width=1\linewidth]{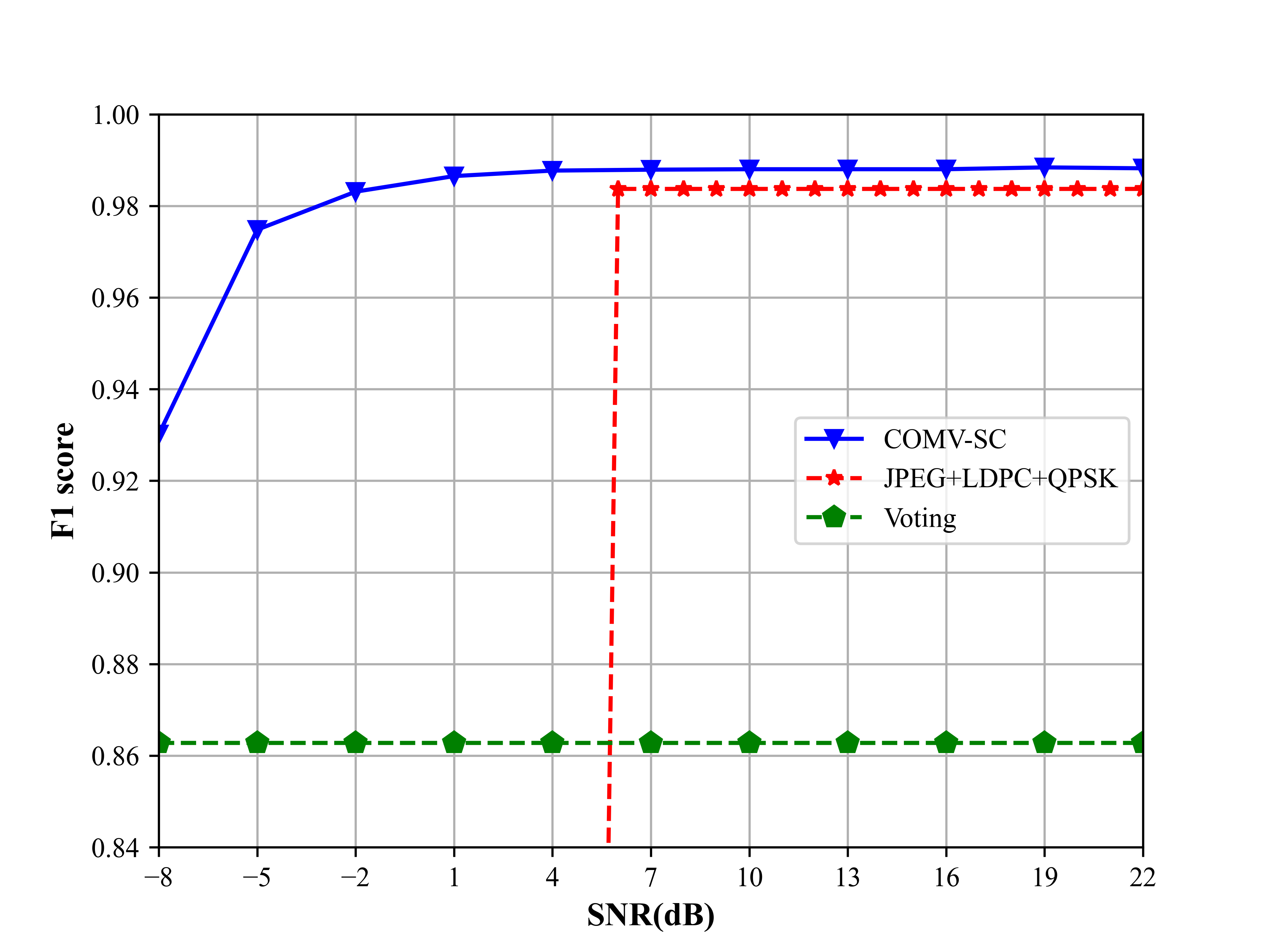}
		\caption{ Comparison of target classification performance with different communication schemes under 45\% occlusion rate.}
		\label{fig_4}
	\end{figure}
	
	\begin{figure}[!t]
		\centering
		\includegraphics[width=1\linewidth]{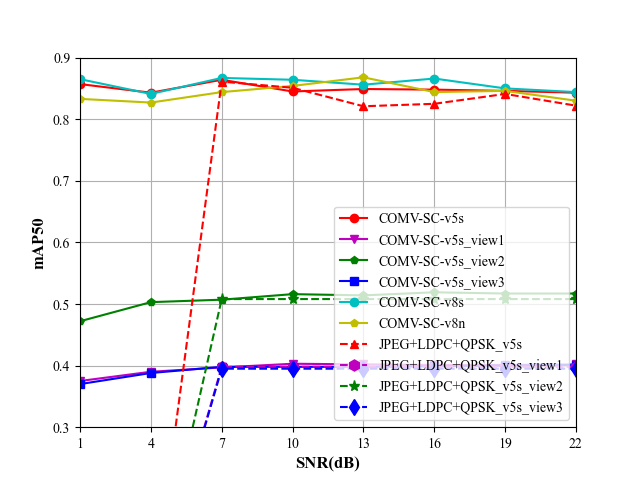}
		\caption{ Comparison of target detection performance between single-vehicle scheme and cooperative multi-vehicle scheme.}
		\label{fig_5}
	\end{figure}

	\subsection{Performance Evaluation}
	Fig. \ref {fig_4} demonstrate the superior performance of the COMV-SC compared to the traditional method and the voting method under different SNRs. Moreover, the occlusion approach in\cite{ref11} is adopted to introduce an occluding object to mask 45\% of target images in the testing set to enhance the difficulty of classification. It can be observed that the COMV-SC achieves better performance with much less transmitted data than the traditional methods. Specifically, the COMV-SC with a compression ratio of 8.4\%, which is about 1/7 of the JPEG+LDPC+QPSK, achieves an F1 score of approximately 0.9880 under moderate and high SNR. Moreover, the COMV-SC overcomes the ``cliff effect” and maintains an F1 score above 0.93 under low SNR. In contrast, the JPEG+LDPC+QPSK fails to operate when the SNR is less than 5dB. Consequently, the COMV-SC can effectively reduce the amount of data to be transmitted while maintaining high performance in autonomous driving.
	
	To validate the superior performance of target detection in the proposed COMV-SC, Fig. \ref {fig_5} further compares the cooperative multi-vehicle scheme with the single-vehicle scheme. Moreover, since the image artefacts and distortions after stitching result in a mismatch with the detection labels of the dataset, we randomly selected 200 clear spliced images at different SNRs and manually labelled them according to the original labels of the dataset. From Fig. \ref {fig_5}, it is evident that cooperative multi-vehicle scheme benefits from a wider perspective, enabling the perception of more targets. The mAP50 metric of the cooperative multi-vehicle scheme is 1.6-2.1 times higher than that of the single-vehicle scheme. Furthermore, compared with YOLOv5s, the superior detection performance of YOLOv8s contributes to marginally improved experimental results, while the lightweight design of YOLOv8n slightly decreases overall performance, demonstrating the stability of the proposed model across different detection algorithms.

    \begin{table}[htbp]  
		\centering  
		\caption{Accuracy results on images from test set: MFNet, RTFNet, and COMV-SC} 
		\label{table:example}  
		\begin{tabular}{cccccccccc}   
			\hline  
			\textbf{Model} & Car & Person & Bike & \makecell[c]{Class \\avg.} & \makecell[c]{Mean \\IoU} \\   
			\hline  
			MFNet & 75.56 & 67.80 & 56.05 & 45.04 & 39.17 \\  
			\hline  
			RTFNet & 91.50 & 76.79 & 73.12 & 61.70 & 49.83 \\
			\hline
			\makecell[c]{COMV-SC \\(0dB)} & 89.06 & 72.22 & 67.63 & 57.56 & 47.12 \\
			\hline
			\makecell[c]{COMV-SC \\(8dB)} & 90.56 & 74.93 & 70.63  & 60.39 & 47.86 \\
			\hline  
			\makecell[c]{COMV-SC \\(15dB)} & 90.76 & 75.42 & 71.18 & 60.92 & 47.96 \\
			\hline  
		\end{tabular}  
	\end{table}  

    As shown in Table 1, the semantic segmentation performance of the COMV-SC are evaluated under varying SNR conditions. Notably, even at an exceedingly low compression rate of 6.25\%, the COMV-SC demonstrates superior performance than the MFNet. When compared with the RTFNet, the proposed system exhibits slightly lower values in terms of two metrics. Considering that the MFNet and RTFNet don not involve transmission and are not affected by channel noise, the COMV-SC is demonstrated to be effective in multi-modal data transmission with greatly reduced data volume. Moreover, these numerical results further underscore the robustness of the proposed system in the presence of channel noise.
    
	The computational complexity of the proposed COMV-SC is evaluated by measuring the average runtime for encoding and decoding one image on a server equipped with one 2.20GHz Intel Xeon Gold 5220R CPU and an NVIDIA RTX A6000 GPU. The JSCC network achieved an average runtime of 8.78ms per image with GPU implementation, significantly outperforming the 261.59ms for JPEG encoding and decoding, without considering the additional time for the compressed bitstream with a channel code. Moreover, the lower compression rate resulted in significantly decreased data transmission time at a fixed transmission rate with less computing times.

	\section{Open Research Directions and Challenges}

	The proposed system represents an initial work to integrate semantics into edge intelligence-enabled autonomous driving for collaborative communications. Substantial further researches are required in the following areas:
	
	\subsubsection{Semantic-aware Resource Allocation}
	
	Enabling resource sharing among multiple users with diverse resources can further enhance the communication efficiency, which however is not considered in the proposed system. How to determine the optimal semantic-aware resource allocation policy in complex scenarios where the transmitted data vary in size and semantic-importance level is a crucial issue.
	
	\subsubsection{Knowledge Base and  Communication Security}
	In autonomous driving,  the construction and updating of the knowledge base to achieve efficient global knowledge sharing among multiple users is a critical challenge that requires innovative solutions. Moreover, it is imperative to conduct further research on developing an effective security policy related to the knowledge base in the proposed system.

	\subsubsection{Joint Communication, Sensing and Computation}
	In the proposed system, the communication, sensing, and intelligent computation functions are separated. However, joint communication, sensing and computation can significantly enhance resource utilization, improve throughput and sensing accuracy, and minimize information interaction overhead. Effectively integrating these functions of the proposed system is an essential direction for future research.

	\section{Conclusions}
	In this article, we proposed a unified multi-user semantic communication system for multiple tasks to support the implementation of edge intelligence enabled autonomous driving system. The proposed system aims to reduce the data volume transmitted in autonomous driving tasks by fusing multi-modal data from multiple users to make more effective decisions with edge servers. The effectiveness of the proposed system is demonstrated through a case study. Experimental results demonstrate that the amount of transmission data can be greatly reduced compared to traditional transmission methods. Moreover, the proposed system efficiently integrates multi-user data, leading to more comprehensive task results for the multiple tasks. Furthermore, we discussed the challenges and open research issues in the proposed system.

	\bibliographystyle{IEEEtran}
	\bibliography{20231229.bib}
	
\end{document}